\documentclass[
superscriptaddress,
reprint,
 amsmath,amssymb,
 aps, physrev,
]{revtex4-2}

\usepackage[T1]{fontenc}
\usepackage{graphicx}
\usepackage{dcolumn}
\usepackage{bm}
\usepackage{siunitx}
\renewcommand

\begin{document}

\title{\textbf{Valley polarization of moiré interlayer exciton complexes driven by many-body interactions} 
}%

\author{Adrian O. Paulus}
\affiliation{%
Walter Schottky Institute and School of Natural Sciences, Technical University of Munich, 85748 Garching, Germany
}%
\affiliation{
 Munich Center for Quantum Science and Technology (MCQST), Schellingstr. 4, 80799 München, Germany
}%

\author{Mikołaj J. Metelski}%
\affiliation{%
Walter Schottky Institute and School of Natural Sciences, Technical University of Munich, 85748 Garching, Germany
}%
\affiliation{
 Munich Center for Quantum Science and Technology (MCQST), Schellingstr. 4, 80799 München, Germany
}%
\author{Alain Dijkstra}%
\affiliation{%
Walter Schottky Institute and School of Natural Sciences, Technical University of Munich, 85748 Garching, Germany
}%
\affiliation{
 Munich Center for Quantum Science and Technology (MCQST), Schellingstr. 4, 80799 München, Germany
}%
\author{Kenji Watanabe}
\affiliation{%
 Research Center for Electronic and Optical Materials,
National Institute for Materials Science, 1-1 Namiki, Tsukuba 305-0044, Japan
}%

\author{Takashi Taniguchi}
\affiliation{%
 Research Center for Materials Nanoarchitectonics,
National Institute for Materials Science, 1-1 Namiki, Tsukuba 305-0044, Japan
}%

\author{Nathan P. Wilson}
\email{Contact author: nathan.wilson@tum.de}
\affiliation{%
Walter Schottky Institute and School of Natural Sciences, Technical University of Munich, 85748 Garching, Germany
}%
\affiliation{
 Munich Center for Quantum Science and Technology (MCQST), Schellingstr. 4, 80799 München, Germany
}%

\author{Jonathan J. Finley}
\affiliation{%
Walter Schottky Institute and School of Natural Sciences, Technical University of Munich, 85748 Garching, Germany
}%
\affiliation{
 Munich Center for Quantum Science and Technology (MCQST), Schellingstr. 4, 80799 München, Germany
}%

\date{\today}

\begin{abstract}
Localized interlayer excitons (IX) in moiré transition metal dichalcogenide heterostructures can both probe and participate in many-body states hosted by the moiré superlattice. 
When the IX density is small compared to the moiré lattice density, the formation of incompressible charge crystals at fractional electronic moiré fillings modifies exciton-charge scattering, leading to enhanced lifetimes in photoluminescence (PL) measurements. 
At high IX densities, the exciton dynamics are altered by the emergence of an excitonic Mott insulator and the formation of doubly-occupied sites (IXX). 
Here, we investigate the IX PL lifetime and valley polarization in an R-type $\mathrm{WSe_2}$/$\mathrm{WS_2}$ bilayer across a wide range of IX and charge densities.
While previous studies reported a decrease of polarization in time-integrated measurements in charge-incompressible phases, our results show that this arises not from enhanced intervalley scattering, but from a dilution of the valley polarization by the dramatic enhancement of IX lifetimes.
At high excitation densities, we probe the dynamics of the IXX and show that despite the nominal antiparallel valley configuration of the two constituent excitons, a strong, anomalous valley polarization develops as the IXX population decays. 
Our results shed light on the complex exciton and valley dynamics of IX and demonstrate that they are strongly modified by the rich many-body physics of moiré heterobilayers.
\end{abstract}

\maketitle

\section{\label{sec:level1}Introduction}
The moiré pattern in heterobilayers of $\mathrm{WSe_2}$ and $\mathrm{WS_2}$ can host both localized charges and interlayer excitons (IX), while promoting correlations due to intra- and inter-species interactions.
Depending on the densities of charges and IX in the lattice, IX can either be considered a passive probe of the charge order, or participate in the many-body phase. 
Coulomb repulsion among charges in the moiré lattice leads to the formation of correlated insulators at fractional and integer lattice fillings~\cite{tang_simulation_2020,xu_correlated_2020,regan_mott_2020,jin_stripe_2021,li_imaging_2021}, 
which can be detected using the IX as a spectroscopic sensor~\cite{miao_strong_2021}.
Interactions between IX are dominated by repulsive dipole-dipole interactions, which result in an energetic ladder for multiply-occupied sites~\cite{park_dipole_2023} or the formation of excitonic crystals at unit lattice filling~\cite{xiong_correlated_2023,gao_excitonic_2024,xiong_tunable_2024,deng_frozen_2025}.
Furthermore, the valley polarization and optical selection rule of IX~\cite{rivera_valley-polarized_2016,rivera_interlayer_2018} can probe spin and valley correlations in the many-body state~\cite{xiong_tunable_2024}. 
However, forming and studying correlated phases with excitons is complicated by the continuous excitation and recombination of excitons over nanosecond to microsecond timescales~\cite{rivera_observation_2015,tugen_optical_2025}. 
As such, correlated phases of excitons are either transient (for pulsed excitation) or non-equilibrium (for continuous excitation).
Moreover, the recombination and scattering dynamics of the IX are significantly influenced by the electronic moiré lattice filling~\cite{miao_strong_2021,liu_excitonic_2021,wang_intercell_2023,tan_layer-dependent_2023,tan_enhanced_2025,upadhyay_giant_2026}.
Time-resolved measurements provide detailed information about how the excitonic lattice population interacts both with itself and the electronic lattice as the exciton population relaxes and then decays.
Such investigations are key for realizing and understanding possible collective dynamics~\cite{kumlin_superradiance_2024,devenica_collective_2026,Brotons-Gisbert2024} or  emergent topological effects~\cite{huang_collective_2024} in excitonic systems.
\par\medskip

The transient nature of the IX and the complex many-body interactions they are subject to motivate this study. 
Here, we probe how the IX population and its valley polarization evolve following injection by a picosecond optical pulse in the presence of correlated electronic states and under different excitation densities. 
We observe enhanced IX lifetimes at charge neutrality and fractional electronic moiré lattice fillings which realize correlated charge insulators, accompanied by a decrease in the time-averaged circular polarization of the emission.
Analysis reveals that IX intervalley scattering is surprisingly insensitive to correlated charge phases, but is nevertheless dependent on overall charge dopant density.
Curiously, this effect depends strongly on the density of injected IX up to the excitonic Mott gap, where excitonic many-body interactions dominate and dynamics are governed by the formation and recombination of doubly-occupied sites (IXX).
The IXX quickly becomes unpolarized following excitation due to the preferred singlet valley pseudospin arrangement of the constituent excitons~\cite{yu_moire_2017,wilson_excitons_2021,wu_highly_2025}, but develops a strong anomalous valley polarization after $\approx$\SI{30}{\nano\second} in the presence of a valley-polarized IX reservoir.
Our results highlight the important role of dynamics in understanding both exciton-charge and exciton-exciton many-body interactions.

\section{Device characterization}
\begin{figure*}[htbp]
    \centering
    \includegraphics[width=\textwidth]{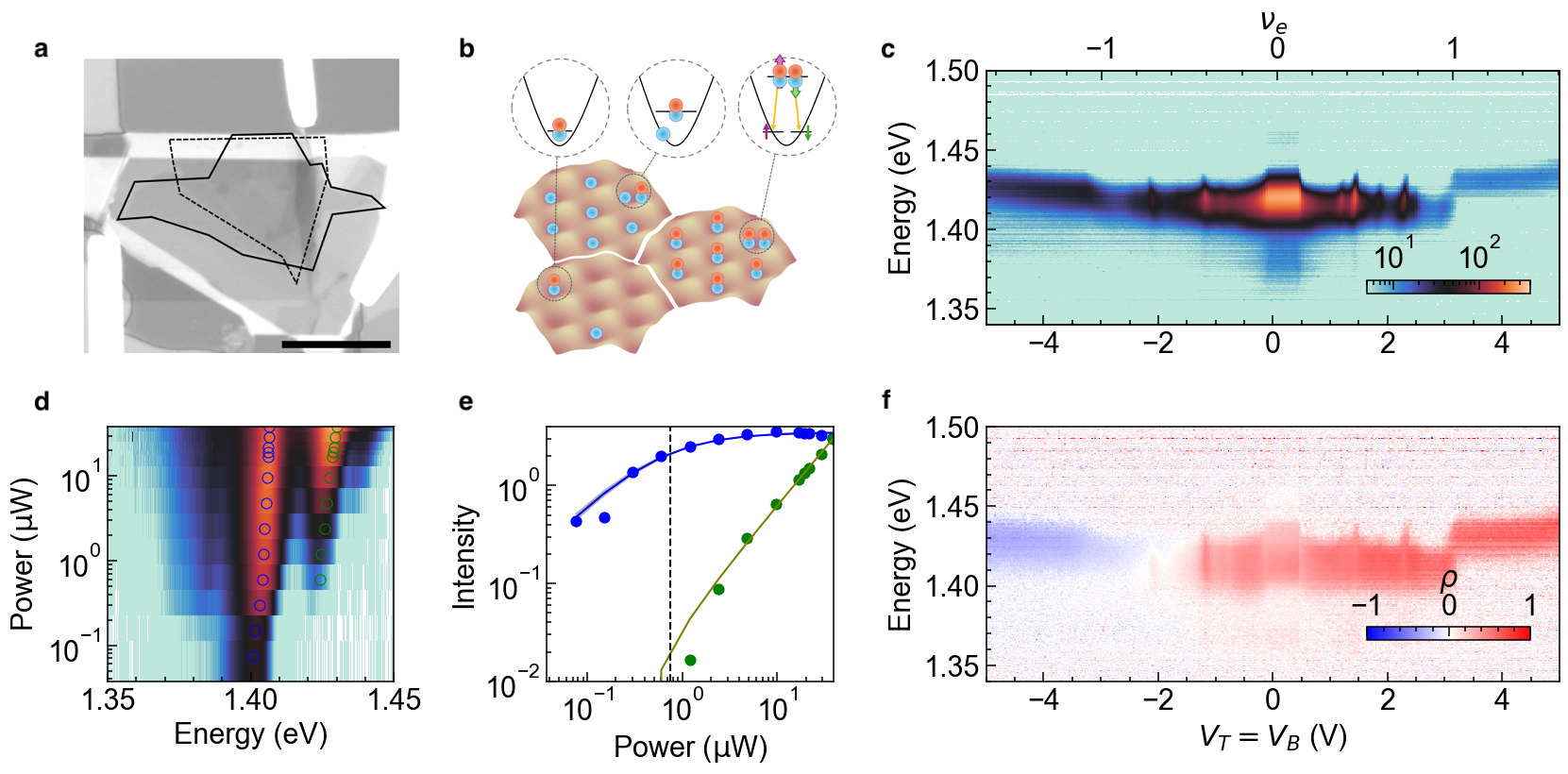}
    \caption{\textbf{Localization of electrons and IX in the moiré lattice. a,} Optical micrograph of device, with $\mathrm{WSe_2}$ ($\mathrm{WS_2}$) outlined in a black (dashed) line. The scale bar corresponds to \SI{20}{\micro\meter}. 
    \textbf{b,} Charges and IX can be localized by the moiré potential. When either $\nu_e > 1$ or $\nu_{\mathrm{IX}} > 1$, particles are forced to cohabit lattice sites and repel one another. 
    \textbf{c,} Doping-dependent CW PL of IX at \SI{750}{\nano\watt} ($\nu_{\mathrm{IX}} \approx 1$) with clear signatures of correlated states. 
    \textbf{d,} IX are filled in discrete steps by modulating excitation power of CW laser. 
    The circular dots represent peak centers extracted from fitting Gaussian line profiles.
    \textbf{e,} IX intensity saturates as soon as every lattice site is occupied, as is indicated by dashed vertical line. At the same time, IXX starts appearing. 
    \textbf{f,} Degree of polarization of IX in CW at \SI{7.5}{\micro\watt} as a function of electronic lattice filling, confirming R-type configuration of device.}
    \label{fig:fig1}
\end{figure*}

We study IX in an angle-aligned dual-gated R-stacked heterobilayer of $\mathrm{WSe_2}$/$\mathrm{WS_2}$ using both pulsed and continuous wave (CW) excitation.
Figure \ref{fig:fig1}a shows an optical image of the device.
Our laser excites intralayer excitons near-resonant to $\mathrm{WSe_2}$, at \SI{1.765}{\eV}, with a pulse width of \SI{20}{\pico\second} in pulsed operation.
We study the photoluminescence (PL) of the IX using time-integrated and time-resolved spectroscopy at a sample temperature of \SI{5}{\kelvin} unless stated otherwise.
\par\medskip
By tuning gate voltages or modulating excitation power, we control moiré lattice occupancy for charges and excitons  \cite{tang_simulation_2020,brotons-gisbert_moire-trapped_2021,park_dipole_2023,xiong_correlated_2023}, respectively, as depicted in Figure \ref{fig:fig1}b.
There, blue circles represent charges and bound pairs of orange and blue circles represent IX.
We quantify the average density of particles in the moiré lattice  via the electron and IX moiré filling factors, $\nu_e$ and $\nu_{\mathrm{IX}}$, respectively.
As we tune $\nu_e$ and $\nu_{\mathrm{IX}}$, the IX spectrum reacts to the changing electronic state and moiré band filling.
When sparsely filled, i.e. $\nu_e$ and $\nu_{\mathrm{IX}} < 1$, the IX and charges spread out and lattice sites contain at most one charge or exciton per site.
We depict this in the bottom left hexagon in Figure \ref{fig:fig1}b.
In this regime, emission of the IX is sensitive to electronic ordering that occurs at fractional values of $\nu_e$, which can be exploited to probe charge correlations~\cite{miao_strong_2021,liu_excitonic_2021}.
Once $\nu_e \geq 1$, the IX is forced to share a moiré lattice site with a charge as depicted in the top left hexagon.
Charges and IX on the same site experience repulsive on-site Coulomb interactions.
Similarly, in the right hexagon, as $\nu_{\mathrm{IX}}$ surpasses one, additional IX must occupy the same lattice site as another IX. 
Due to their dipolar nature, the IX are then subject to repulsive Coulomb interactions.
Notably, as $\nu_{\mathrm{IX}}$ reaches one, the direct Coulomb interactions cause the IX to order into an excitonic Mott insulator~\cite{xiong_correlated_2023,xiong_tunable_2024}.
Then, as $\nu_{\mathrm{IX}}$ exceeds one, on doubly occupied sites, the IX prefer the energetically-favorable antiparallel valley pseudospin singlet arrangement due to repulsive exchange interactions~\cite{yu_moire_2017,xiong_tunable_2024}.
\par\medskip
We begin probing the device using CW excitation at \SI{750}{\nano\watt} and observe typical gate-dependent PL in Figure \ref{fig:fig1}c.
As we sweep over the gate voltages $V_G$, IX PL is enhanced at specific electronic fillings.
These enhancements are signatures of correlated states at fractional values of $\nu_e$, consistent with previous reports ~\cite{miao_strong_2021,liu_excitonic_2021}.
We identify the Mott insulator from the sharp energy jump of the IX around $V_G = \SI{3.2}{\volt}$ and use it to calibrate our electronic moiré lattice filling, as described in previous works~\cite{tang_simulation_2020,xu_correlated_2020} (see appendix for details).
\par\medskip
We continue by analyzing the power-dependent behavior of the IX  under CW excitation at $\nu_e=0$ in Figure \ref{fig:fig1}d.
With increasing power, IX intensity increases and its central energy blueshifts.
At a threshold power of approximately \SI{750}{\nano\watt}, a second peak corresponding to double IX occupancy of a moiré lattice site, denoted IXX, appears 30 meV blueshifted from the IX~\cite{park_dipole_2023}.
As the power is further increased, the intensity $I_{\mathrm{IX}}$ of IX appears to saturate while that of the IXX $I_{\mathrm{IXX}}$ increases with a slightly superlinear power law $I_{\mathrm{IXX}}(P) \propto P^{\alpha}$ with $\alpha \approx 1.2$.
We confirm this behavior by extracting power dependent intensity in Figure \ref{fig:fig1}e.
The brightness of IX increases with power up to the onset of IXX.
This behavior is characteristic for IX localized in strong moiré potentials ~\cite{park_dipole_2023,xiong_correlated_2023}.
Accordingly, we can establish a low and high power regime based on this threshold power. 
\par\medskip
To verify the stacking order, we plot the degree of circular polarization of the IX as a function of the electronic lattice filling in Figure \ref{fig:fig1}f, and observe characteristic spectral fingerprints for R-type alignment~\cite{wang_intercell_2023,kim_correlation-driven_2024}.
In particular, around charge neutrality, the IX is co-circularly polarized.
At higher electron doping, the degree of polarization increases, and once doped past the electronic Mott insulator at $\nu_e = 1$, the degree of polarization approaches unity.
Upon hole doping the lattice, the polarization switches abruptly at a threshold moiré lattice filling~\cite{wang_intercell_2023}.
We note that the degree of polarization is slightly reduced at charge correlated insulators as well as at charge neutrality.
\par\medskip
This reduction in circular polarization is intriguing: while previous investigations hypothesize this to be a consequence of enhanced intervalley scattering caused by a reduction of screening~\cite{liu_excitonic_2021}, other effects may also play a role.
For example, charge crystallization could impact the exciton decay dynamics by enhancing or closing off additional recombination pathways.

\par\medskip
\begin{figure}[htbp]
    \centering
    \includegraphics[width=\linewidth]{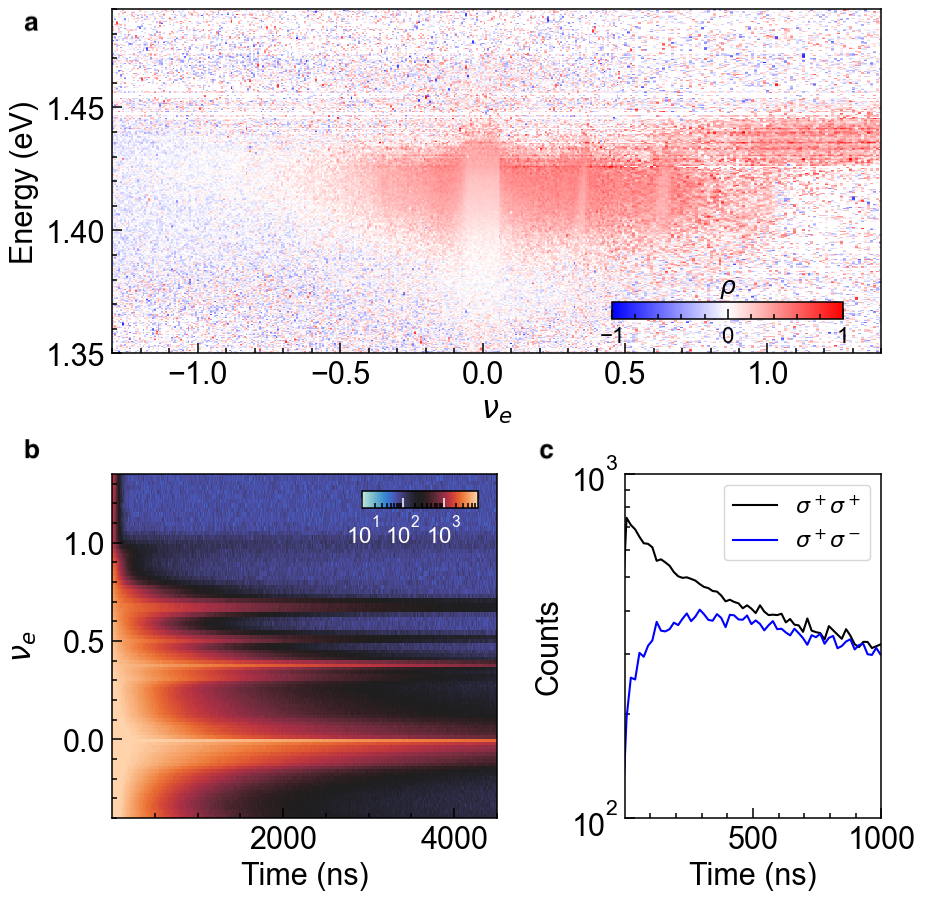}
    \caption{\textbf{Polarization dependence of IX emission under pulsed excitation. a,} Degree of polarization dependent as a function of electronic lattice filling with \SI{200}{\kilo\hertz} pulsed excitation at \SI{1.5}{\micro\watt}. Degree of polarization decreases at correlated electronic states and charge neutrality. 
    \textbf{b,} Time-resolved PL of the IX at \SI{1.5}{\micro\watt}. \textbf{c,} Polarization resolved transients of the IX at $\nu_e=0$ and \SI{300}{\nano\watt}.}
    \label{fig:fig2}
\end{figure}

To distinguish between these different types of effects, we directly probe the valley and PL dynamics of the IX in the presence of a correlated charge insulators. The following measurements were performed with pulsed excitation at a repetition rate of \SI{200}{\kilo\hertz} unless specified otherwise.
We plot the degree of circular polarization of the IX as a function of the electronic lattice filling in Figure \ref{fig:fig2}a, this time with pulsed excitation.
We recover similar polarization behavior to the CW case, namely co-circularly polarized emission whenever $\nu_e > 0$, and a reversal of the polarization upon hole doping.
Moreover, the degree of circular polarization in our data clearly reduces at correlated fractional charge fillings.
\section{Valley dynamics across electronic filling regimes}
To understand the effect of the correlated electronic states, we performed time correlated single photon counting to track the PL dynamics as a function of gate voltage and polarization at \SI{1.5}{\micro\watt}, shown in Figure \ref{fig:fig2}b.
We select the IX emission spectrally with a monochromator before detection with a single photon avalanche detector (see appendix for details).
Time-integrated intensity and lifetimes are notably enhanced at correlated electronic states and at $\nu_e = 0$, i.e. in charge incompressible states~\cite{tan_layer-dependent_2023,yan_anomalously_2025,upadhyay_giant_2026}.
In agreement with previous reports~\cite{yan_anomalously_2025}, we attribute prolonged lifetimes to the crystallization of the electrons into a lattice, reducing exciton-electron scattering, thereby suppressing non-radiative decay channels.
This is supported by temperature dependent decay dynamics, which show that with increasing temperature, the correlated states begin to melt and the IX lifetime and integrated intensity decrease (see extended data Figure \ref{fig:ext_fig1}).
Once melted, emission at the correlated fractional fillings is indistinguishable from uncorrelated fillings in both intensity and lifetime.
\par\medskip
To understand the connection between the correlated electronic states and the observed decrease in valley polarization in time-integrated measurements, we resolve the valley dynamics by recording polarization dependent PL transients.
Taking $\nu_e = 0$ (Figure \ref{fig:fig2}c), the co- and cross-polarized channels evolve differently:
In the co-circularly polarized channel, emission peaks immediately after the laser pulse, and decreases monotonically over time.
This stands in contrast to the cross-circularly polarized channel, where the initial signal is much lower, increasing over tens of nanoseconds until peaking at \SI{300}{\nano\second}, before gradually decreasing again.
\par\medskip
This behavior is a result of the valley lifetime at $\nu_e=0$ being much shorter than the PL lifetime.
This gives the valley polarization enough time to decay before the IX population recombines.
At least for $\nu_e=0$ this means that a significant portion of the emitted light will be unpolarized.
Of course, if the valley lifetime becomes comparable to or longer than the PL lifetime, emission will occur before valley polarization is lost, resulting in higher recorded PL circular polarization in time-integrated measurements.
\par
\begin{figure*}[htbp]
    \centering
    \includegraphics[width=\textwidth]{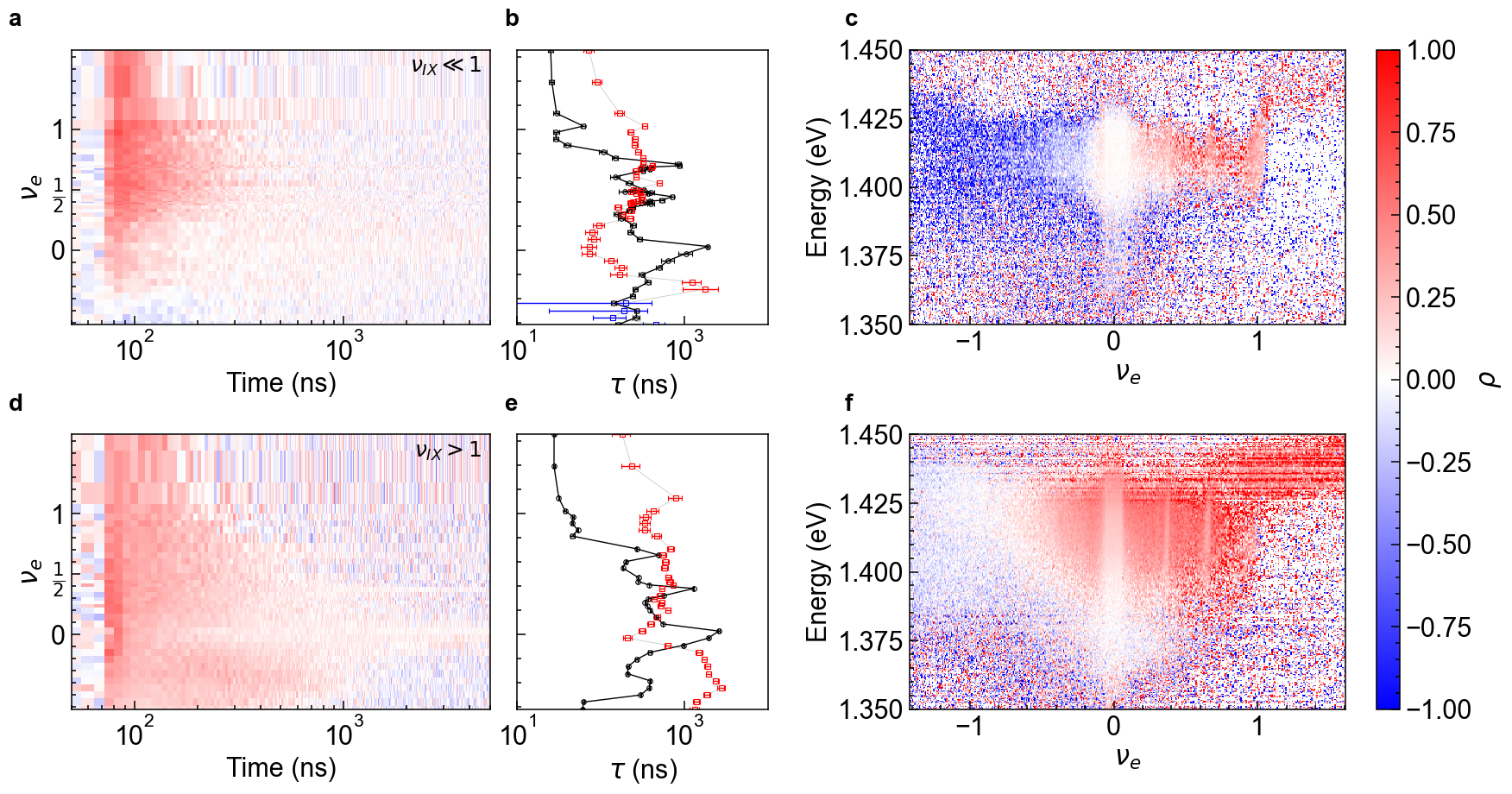}
    \caption{\textbf{Gate control of IX polarization dynamics.
    a,} Time resolved degree of polarization as a function of electronic lattice filling under pulsed excitation at \SI{37.5}{\nano \watt} ($\nu_{\mathrm{IX}} \ll 1$), and \textbf{d,} at \SI{1.5}{\micro \watt} ($\nu_{\mathrm{IX}} > 1$).
    Polarization responds to changing overall charge density, but remains relatively unaffected by correlated electronic states.
    \textbf{b and e,} Fits to PL lifetime (black) and valley lifetimes at low and high excitation power respectively.
    Red color indicates co-circular polarization, blue cross-circular polarization.
    PL lifetimes respond strongly to correlated fillings, while valley lifetimes are weakly affected.
    \textbf{c and f,} Degree of polarization in PL spectra under the same excitation conditions as in a and d respectively. The observed behavior mirrors time resolved data, however, at commensurate fillings, degree of polarization is masked due to long PL lifetimes.}
    \label{fig:fig3}
\end{figure*}

\medskip
With this in mind, we probe the valley dynamics as a function of filling factor, at low pump power (\SI{37.5}{\nano \watt}, $\nu_{\mathrm{IX}} \ll 1$).
Figure \ref{fig:fig3}a shows the degree of circular polarization of the PL as a function of time.
Three regions are identified:
Region 1, where $\nu_e>1$, region 2, where $0<\nu_e<1$, and region 3, where $-1<\nu_e<0$.
In region 1, we observe near unity initial valley polarization coupled with short PL lifetimes (see Figure \ref{fig:fig2}b) .
In region 2, valley lifetimes depend strongly on the charge density.
Shortly after excitation, polarization is  strong and valley lifetimes are generally long. 
The valley lifetime is longest around $\nu_e=0.5$ and falls off towards $\nu_e=1$ and $\nu_e=0$.
Despite the strong doping dependence in region 2, only a weak change in valley lifetime, if any, is visible at correlated fractional charge insulators.
In region 3, upon hole doping, the valley lifetime briefly increases again, before the polarization drops.
Adding more holes causes emission to become cross-polarized.
\par\medskip
We extract valley and PL lifetimes in Figure \ref{fig:fig3}b (for more details see Appendix A (methods); select polarization resolved transients can be found in appendix Figure \ref{fig:ext_fig4}).
The black points indicate PL lifetime, while red (blue) points indicate valley lifetime with co(cross)-circular emission.
Valley lifetimes show at best a weak enhancement by correlated electronic states.
At the same time, in incompressible states, the PL lifetime is \textit{significantly} enhanced by up to an order of magnitude, far exceeding the valley lifetime in some cases.
This can explain the reduction in polarization observed in time-integrated measurements shown previously in Figure \ref{fig:fig2}a, and repeated in Figure \ref{fig:fig3}c under the same experimental conditions as Figure \ref{fig:fig3}a.
In the presence of incompressible electronic states, PL lifetimes are sufficiently long for the IX to fully depolarize, and a substantial amount of signal originates from the unpolarized emission tail. 
From these observations, we conclude that the enhanced PL lifetimes simply \textit{dilute} the polarization of the IX, while the incompressible states have little direct effect on the valley dynamics.
\par\medskip
We repeat these time resolved measurements in Figure \ref{fig:fig3}d at \SI{1.5}{\micro\watt},  increasing IX density to $\nu_{\mathrm{IX}} > 1$.
In comparison to the lower power regime (Figure \ref{fig:fig3}a), the valley lifetime is generally longer, although away from charge neutrality, the initial polarization appears to be weaker.
Tracking degree of polarization versus charge lattice filling $\nu_e$, valley lifetimes on the hole and electron-doped sides now appear almost symmetric around the point of charge neutrality, with the co-circularly polarized emission extending well into the hole side.
At charge neutrality, valley polarization remains inhibited even though PL lifetimes are long.
\par\medskip
As before, we extract these lifetimes in Figure \ref{fig:fig3}e.
As in the lower power regime, valley lifetimes  are mostly unaffected by the presence of correlated states, while the PL lifetimes are greatly enhanced, leading to the same observation of reduced polarization in time-integrated measurements (Figure \ref{fig:fig3}f).
Nevertheless, at all filling factors, compared to the low power regime, valley lifetimes are longer by roughly an order of magnitude, while the PL lifetimes are only slightly increased, leading to the overall larger polarization observed for high excitation power in the time-integrated measurement. 
Additionally, the reversal of polarization upon hole doping appears to depend strongly on the exciton density, pointing towards possible many-exciton effects influencing polarization dynamics.

\section{Exciton-exciton interactions driving valley dynamics}
Simple arguments about the valley and PL lifetimes capture the observed change in polarization in charge incompressible states, but other aspects of the data suggest the involvement of many-body interactions in shaping the IX dynamics.
In particular, the cross-polarization in the hole doped regime and the initial polarization at short times after excitation appear to depend on excitation density, implying the influence of exciton-exciton interaction. 
In Figures \ref{fig:fig3}d-f, a sufficiently large excitation power (\SI{1.5}{\micro \watt}) is used, such that $\nu_{\mathrm{IX}} > 1$.
There, dipole-dipole interactions lead to strong correlations among the IX, which can lead to the formation of an excitonic Mott insulator~\cite{xiong_correlated_2023,lian_valley-polarized_2024}.
Even absent a global many-body state,  moiré lattice sites occupied by two IX (doublons) are subject to strong Coulomb interactions, manifesting as the spectrally blueshifted exciton complex IXX~\cite{park_dipole_2023}.
\par
\begin{figure*}[htbp]
    \centering
    \includegraphics[width=\textwidth]{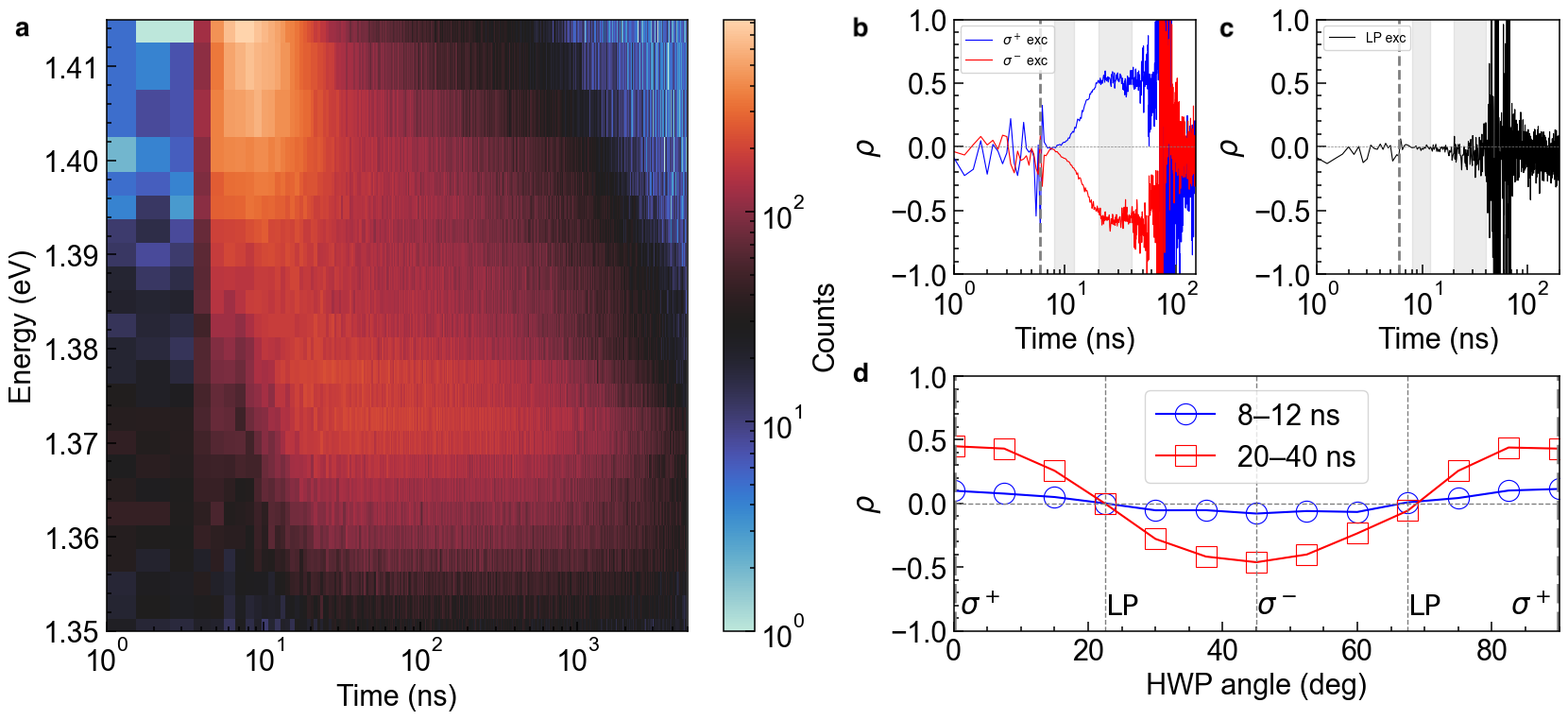}
    \caption{\textbf{Energy resolved dynamics of IX and IXX. 
    a,} Energy and time resolved emission of IX and IXX. \textbf{b,} Degree of circular polarization of IXX under $\sigma^+$ (blue) and $\sigma^-$ (red), or \textbf{c,} linear polarization.
    \textbf{d,} Degree of polarization under varying excitation polarization at different times after excitation, corresponding to different moiré lattice fillings $\nu_{\mathrm{IX}}$.}
    \label{fig:fig4}
\end{figure*}

\medskip
We investigate the dynamics of the IXX in Figure \ref{fig:fig4} at $\nu_{\mathrm{IX}} > 1$ and $\nu_{e}  = 0$.
In Figure \ref{fig:fig4}a, we resolve the temporal evolution of the PL spectrum.
Similar to the time-integrated measurements in Figure \ref{fig:fig1}c, we resolve two emission peaks belonging to the IX and IXX respectively.
IXX emission begins immediately after excitation, while IX emission only begins once IXX has decayed significantly, consistent with previous reports~\cite{deng_frozen_2025}.
This indicates that the number of doubly-occupied lattice sites is initially much larger than the number of singly-occupied sites.
Because the IXX decays on \SI{10}{\nano\second} timescales, two orders of magnitude faster than the IX at $\nu_{e}  = 0$, the lattice occupancy quickly decays towards $\nu_{\mathrm{IX}}=1$.
\par\medskip
Next, we turn to the valley polarization of the IXX.
Due to repulsive exchange interactions, the ground state of the IXX is expected to be the valley pseudospin singlet configuration~\cite{park_dipole_2023} , in which the two constituent excitons reside in opposite valleys. 
Absent any breaking of time reversal symmetry, the relaxation of an IXX to an IX by recombination of an exciton from either valley should be equally likely.
Consequently, emission from the IXX has been shown to be unpolarized~\cite{park_dipole_2023,xiong_tunable_2024,jiang_tuning_2025}. 
However, as we will demonstrate, their emission can become strongly polarized in the presence of a valley-polarized population of IX, which explicitly breaks time reversal symmetry.

We show time-resolved degree of polarization under $\sigma^+$ (blue) and $\sigma^-$ (red) excitation in Figure \ref{fig:fig4}b.
Within the first nanosecond after excitation (gray dashed line), the emission appears to be briefly polarized before undergoing rapid depolarization.
For around \SI{10}{\nano\second} after excitation (indicated by the leftmost gray shaded box), emission is largely unpolarized (see appendix Figure \ref{fig:ext_fig5}b).
It is in this time span that most IXX emission occurs and IX emission is weak (see Figure \ref{fig:fig4}a), indicating that most sites are doubly occupied.
In this case, IXX emission is occurring when the IXX is unpolarized, matching expectations from time integrated experiments where the degree of circular polarization for IXX was low~\cite{park_dipole_2023,lian_valley-polarized_2024}.
Surprisingly, as the IXX population depletes, i.e. as $\nu_{\mathrm{IX}} \to 1$, the degree of polarization of the IXX rises dramatically and peaks at around 0.6 (rightmost gray shaded box), before the IXX signal fades entirely.
As the IXX signal fades away, the degree of polarization only decreases slightly, staying above $\sim$0.5 until the signal is obscured by noise and the dark counts of the detector. 
The preferential emission of an IXX from the majority valley suggests that an IX of the \textit{minority} valley is left behind.
In Figure \ref{fig:fig4}c, we repeat these measurements under linearly polarized excitation and do not observe any resurgent (circular) polarization.
\par\medskip
Finally, in Figure \ref{fig:fig4}d, we characterize the polarization behavior of the IXX further by sweeping over the excitation polarization.
We distinguish between two time periods, as indicated by the gray areas in Figure \ref{fig:fig4}b, where the moiré lattice filling is $\nu_{\mathrm{IX}} > 1$ or $\nu_{\mathrm{IX}} \approx 1$ respectively.
Angles \ang{0} and \ang{45} correspond to excitation under $\sigma^+$ ($\sigma^-$) polarization in Figure \ref{fig:fig4}b respectively, and \ang{22.5} to linear polarization in Figure \ref{fig:fig4}c.
The net degree of polarization is low when $\nu_{\mathrm{IX}} > 1$, independent of excitation.
However, as we switch between circular and linear polarized excitation, when $\nu_{\mathrm{IX}} \approx 1$, degree of circular polarization responds sensitively to excitation, and is largest under $\sigma^+$ ($\sigma^-$) excitation.
\par\medskip
Under the same excitation conditions with circularly polarized light, as in Figure \ref{fig:fig4}d, the IX emission has a degree of circular polarization of around 0.3 during the time span when the IXX emission becomes strongly polarized.
Crucially, the degree of polarization of the IXX is nearly double that of the IX (see appendix Figure \ref{fig:ext_fig6}), which excludes the possibility that the detected IXX polarization is simply the result of a polarized high energy spectral tail coming from the IX which overlaps with the IXX in energy.
What we observe is therefore a \textit{repolarization} of the IXX.
We can also conclude that the resurgent polarization cannot stem from interactions among the IXX, as emission is completely unpolarized on shorter timescales, indicating that the IXX population has no net \textit{valley} polarization.
Instead, our measurements suggest that the described behavior originates from preferential relaxation of the IXX to leave behind an IX in the minority valley. Since this process breaks time reversal symmetry and the polarized IX population is the only source of symmetry breaking, the preferential relaxation must be caused by interactions with the valley-polarized IX population. This is further confirmed by  the effect disappearing under linear polarization, when the IX population is unpolarized, and it only arises at timescales when the density of singly occupied sites is close to 1 and the number of  doubly occupied sites (which have a valley-singlet configuration) is small.

\section{Discussion and Outlook}
In summary, we have studied the population and polarization dynamics of the IX in two different density regimes. 
When $\nu_{\mathrm{IX}}$ is small, the exciton acts as a minimally invasive sensor of charge order.
There, the IX valley lifetime is sensitive primarily to the total number of charges, but relatively insensitive to the correlated insulators that occur at commensurate fillings of the moiré lattice.
At the same time, the exciton lifetime is greatly enhanced in the presence of correlated insulators, far exceeding the valley lifetime in those states, giving rise to an apparent \textit{decrease} in polarization in time-integrated measurements.
\par\medskip
On the other hand, when $\nu_{\mathrm{IX}}>1$, exciton-exciton interactions appear to play a significant role in polarization dynamics of the IX.
By increasing $\nu_{\mathrm{IX}}$ the valley lifetime is extended by up to an order of magnitude, an effect which likely underlies the observation of increased degree of polarization at high excitation powers~\cite{lian_valley-polarized_2024}.
The creation of doubly occupied sites with strong pumping also gives clear signatures for the role of exciton-exciton interactions. 
The IXX shows signs of circular polarization in the first \si{\nano\second} after excitation, before rapidly depolarizing.
This indicates that the doublons initially consist of two excitons from the majority valley, which quickly relax to the energetically-favorable antiparallel alignment due to valley-dependent exchange interactions~\cite{yu_moire_2017,park_dipole_2023,xiong_tunable_2024}.
Following the depolarization, the IXX unexpectedly \textit{regains} a high degree of circular polarization as the number of doublons becomes small and $\nu_{\mathrm{IX}}$ approaches $1$.
Since the IXX population has no valley polarization at shorter timescales, we conclude that the polarization of the PL at longer timescales arises due to preferential recombination of an exciton from the majority valley (defined by the background polarization of the IX reservoir). 
Since this recombination process leaves behind an IX from the minority valley, it has the effect of reducing the overall valley polarization of the system.
This could indicate that nearest neighbor IX-IX interactions favor the antiparallel alignment, which could arise from exchange interactions or a dynamic correlation mechanism~\cite{nakagawa_dynamical_2020}.
\par\medskip
We have thus demonstrated how comparatively simple time-resolved measurements can be used to capture intriguing many-body dynamics.
Our study of the IX valley dynamics in the presence of correlated charge insulators opens the door to improved microscopic understanding of interactions between the excitonic and electronic lattices \cite{huang_spin-mediated_2023,huang_mott-moire_2023}.
Similarly, the sensitivity of the IXX dynamic polarization to the background polarization of the IX makes it well suited to probe for valley pseudospin ordering which may occur in correlated phases of excitons~\cite{lian_valley-polarized_2024,xiong_tunable_2024,huang_spin-mediated_2023,huang_mott-moire_2023}, or possibly even valley polarization of charges or combinations of the two~\cite{huang_collective_2024,kumlin_superradiance_2024,gu_dipolar_2022,zhang_correlated_2022}.

\begin{acknowledgments}
The authors would like to thank Michael Knap and Fabian Pichler for helpful discussions. 
A.O.P. acknowledges funding by the Bavarian Hightech Agenda within the Munich Quantum Valley doctoral fellowship program. 
J.J.F. gratefully acknowledges the German Science Foundation for funding via MCQST (EXC-2111, project number 390814868), e-conversion (EXC-2089, project number 390776260), SPP-2244 and FI947-8. 
K.W. and T.T. acknowledge support from the CREST (JPMJCR24A5), JST and World Premier International Research Center Initiative (WPI), MEXT, Japan.
\end{acknowledgments}

\section*{Author contributions}
A.O.P. conceived the work, A.O.P. fabricated the devices, A.O.P., M.J.M. and A.D. performed the optical measurements, A.O.P., N.P.W. and J.J.F. analyzed the results, K.W. and T.T. grew bulk hBN crystals, A.O.P., N.P.W. and J.J.F. prepared the manuscript with input from all authors.

\section*{Data availability}
The data sets generated and analyzed during the current study are available upon reasonable request.

\section*{Competing interests}
The authors declare no competing interests.

\appendix

\section{Methods}

\subsection{Device assembly}
All flakes were mechanically exfoliated on silicon wafers capped with \SI{70}{\nano\meter} of $\mathrm{SiO_2}$.
The crystallographic axes of both $\mathrm{WSe_2}$ and $\mathrm{WS_2}$ were identified optically based on straight edges and well-defined corner angles.
The device was assembled using standard dry stacking assembly using polycarbonate stamps~\cite{Wang_One-Dimensional_2013}.
Gates are spaced symmetrically around the heterobilayer separated by $\approx \SI{50}{\nano\meter}$ thick hBN.
The R-type stacking configuration was confirmed by its distinctive gate dependent degree of polarization (see Figure \ref{fig:fig1}f).
The heterobilayer is directly contacted by a few-layer graphite contact flake and electrodes are defined using standard optical lithography techniques.

\subsection{Time-integrated spectroscopy}
All optical measurements were conducted in a closed-cycle optical cryostat (attoDRY800).
We probed the device by focusing either a \SI{633}{\nano \meter} laser (CW), or a \SI{703}{\nano \meter} diode laser (CW or pulsed) onto the sample using a 40X objective ($\mathrm{NA}=0.6$).
PL from the sample was collected by the objective and focused through a diffraction limited pinhole acting as a spatial filter. 
Finally, a CCD captured the spectrum using a 500mm grating spectrograph (Andor Shamrock 500i).
Our \SI{633}{\nano \meter} laser has a diffraction limited spot size of approximately \SI{1}{\micro\meter}, whereas our \SI{703}{\nano \meter} laser has an elliptical spot size intrinsic to the laser head of approximately \SI{2}{\micro\meter}/\SI{3}{\micro\meter} along the minor/major axis, contributing to the different filling factor calibration $\nu_{\mathrm{IX}}$ for the individual lasers.
The spatially extended excitation spot of the \SI{703}{\nano \meter} laser should initialize the moiré lattice more homogeneously over the collected spot size.

\subsection{Time-resolved spectroscopy}
Time-resolved PL measurements were conducted under similar experimental conditions as the time-integrated spectroscopy.
We exclusively used the \SI{703}{\nano \meter} diode laser under pulsed operation (typical repetition rate \SI{200}{\kilo \hertz}) with pulses $<$\SI{70}{\pico\second}.
Spectral resolution in the experiment was accomplished by using the spectrograph with an exit slit to function as a monochromator with tunable bandwidth. 
After passing through the exit slit, light was coupled into a multi-mode fiber attached to a low dark count single photon avalanche detector (Excelitas SPCM-AQRH-16-FC).
The detected photons were counted by a time-tagger (PicoQuant PicoHarp 300).
We characterized the instrument response function as \SI{890}{\pico\second}, the large temporal broadening most likely a result of the timing jitter of the detector.
The spectral resolution of time resolved measurements was measured to be \SI{2}{\nano\meter} by scanning a narrowband laser source across the slit, which is significantly narrower than the expected FWHM of a typical IX.

\subsection{Polarization resolved measurements}
For polarization resolved measurements, linear polarizers and half wave plates on rotating mounts were installed in the excitation and collection paths, while a quarter wave plate was installed in the common path for excitation/collection near the objective. 
The wavelength ranges of the polarizers and wave plates were chosen to be optimized for the IX wavelength, but still performed well at the laser wavelength, reaching a laser extinction ratio of  $\approx 150$.

\subsection{Fitting}
We model the IX dynamics in the system with a two-level rate equation with two degenerate excited states corresponding to the valleys.
Depending on the valley, an IX is either in population $N_{K}$ or $N_{K'}$.
It can scatter between the valleys with rate $\Gamma_{KK'}$, or decay (non)radiatively with rates $\Gamma_{nr}$, $\Gamma_{r}$ respectively.
The population $\mathbf{N} = \begin{pmatrix}
N_{K}\\
N_{K'} 
\end{pmatrix}$ then evolves according to M, with
\begin{equation}
    \frac{d\mathbf{N}}{dt} = \mathbf{M} \mathbf{N}
\end{equation}
\begin{equation}
    \mathbf{M} = \begin{pmatrix}
-(\Gamma_r + \Gamma_{nr} + \Gamma_{KK'}) & \Gamma_{KK'} \\
\Gamma_{KK'} & -(\Gamma_r+ \Gamma_{nr} + \Gamma_{KK'}) 
\end{pmatrix}.
\end{equation}
Some assumptions are made:
There is no valley preference.
This means that the IX population recombines at the same rate independent of valley, non-radiative processes have no preference in valley.
\par\medskip
We extracted PL lifetimes by fitting a biexponential
\begin{equation}
    N = A_1 \exp(-t/\tau_1) + A_2 \exp(-t/\tau_2) + \mathrm{const.}.
\end{equation}
Wherever the fit was overparameterized, we chose a simple exponential 
\begin{equation}
    N = A_1 \exp(-t/\tau_1) + \mathrm{const.},
\end{equation}
as a model instead.
\par\medskip
We calculated time resolved degree of polarization $\rho(t)$ using 
\begin{equation}
    \rho(t) = \frac{N_+ - N_-}{N_+ + N_-  - 2\, \mathrm{dark counts}}.
\end{equation}
Since we assume no valley preference, we directly fit to
$\rho(t)$ with a single exponential to extract valley lifetimes.
The intervalley scattering time is $\tau_{KK'} = 1/\Gamma_{KK'} = 2\,\tau_{\mathrm{fit}}$, and the valley lifetimes reported throughout are $\tau_{KK'}$.

\subsection{Estimation of filling factors}
We approximate $\nu_e$ by using three points in our measurements:
$\nu_e=0, 1/2 , 1$, and assume piecewise-linear spacing.
These points are identified from the known enhancement of IX PL in correlated charge states.
The corresponding filling factors were assigned based on the relative spacing to other correlated insulators (e.g. the symmetry of the $\nu_e=1/3, 2/3$ about $\nu_e=1/2$), and match well with estimates from a parallel plate capacitor model.
We choose three points to account for nonlinearity in filling close to the point of charge neutrality.

\par\medskip
We determine $\nu_{\mathrm{IX}}$ based on power-series measurements, similar to~\cite{xiong_correlated_2023}.
For time-integrated measurements, we use the onset of IXX luminescence to estimate the point $\nu_{\mathrm{IX}}=1$.
For our time-resolved measurements, the PL signature of the IXX is too weak to resolve clearly at low repetition rates.
Therefore, for time resolved measurements, we base our estimation of $\nu_{\mathrm{IX}}=1$ on the saturation of IX emission intensity measured by the SPADs.
We corroborate this estimate, by verifying emission from the IXX, and the delayed emission onset of the IX as reported for example in~\cite{deng_frozen_2025}.

\section{Extended data}
\begin{figure*}[htbp]
    \centering
    \includegraphics[width=\textwidth]{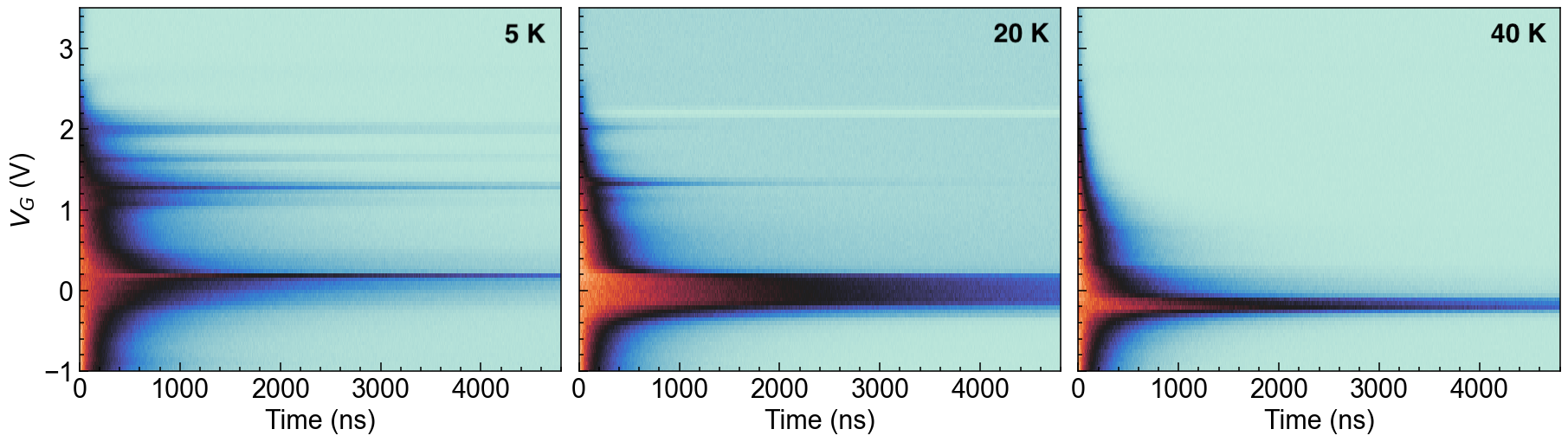}
    \caption{\textbf{IX PL lifetimes at different temperatures.} Time resolved IX PL for temperatures \SI{5}{\kelvin}, \SI{20}{\kelvin}, and, \SI{40}{\kelvin}. As the charge correlated states melt, the extended PL lifetimes at charge correlated states extinguish.}
    \label{fig:ext_fig1}
\end{figure*}

\begin{figure*}[htbp]
    \centering
    \includegraphics[width=\textwidth]{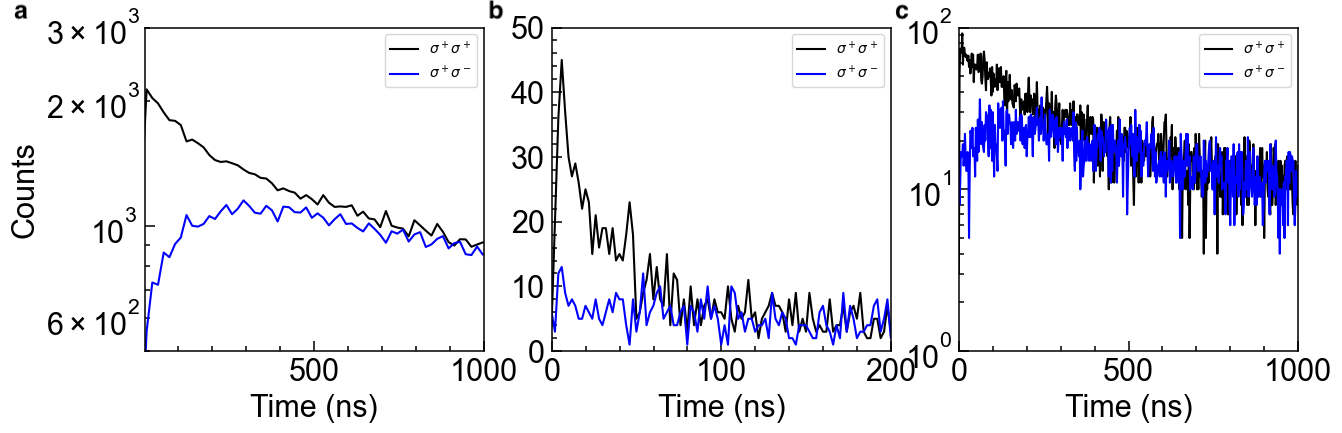}
    \caption{\textbf{Selected polarization resolved traces. a,} Captured at $\nu_e = 0$, and \SI{300}{\nano\watt}. 
    \textbf{b,} Captured at $\nu_e > 1$, and \SI{75}{\nano\watt}. 
    \textbf{c,} Captured at $1/2 < \nu_e < 2/3 $, and \SI{75}{\nano\watt}.}
    \label{fig:ext_fig4}
\end{figure*}

\begin{figure*}[htbp]
    \centering
    \includegraphics[width=\textwidth]{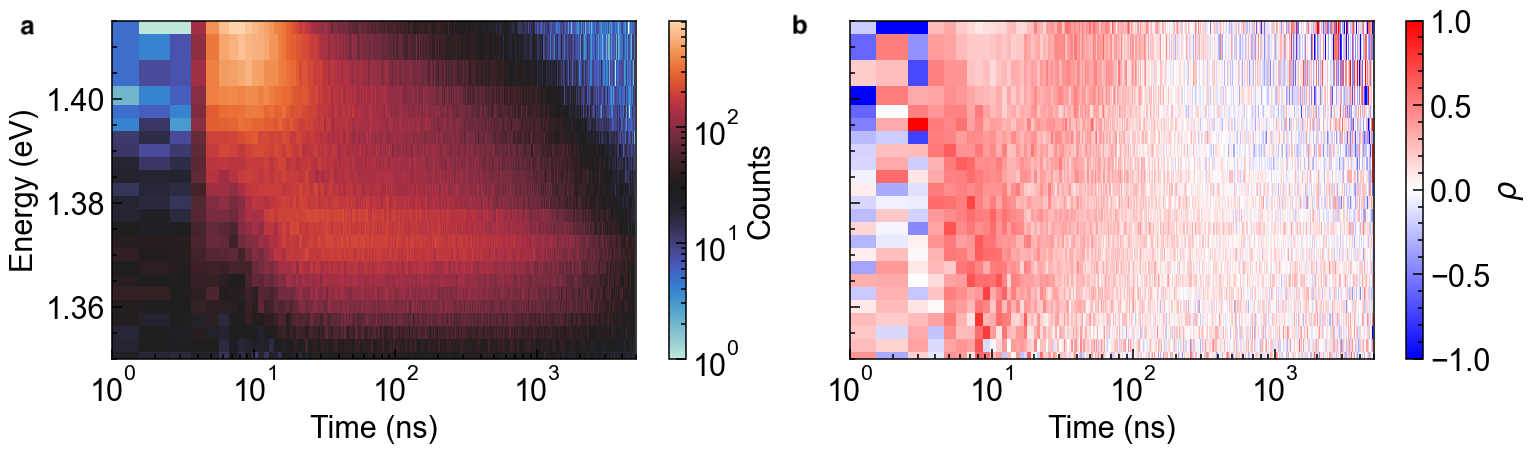}
    \caption{\textbf{Energy resolved IX and IXX emission under $\sigma^+$ excitation. a}, Sum of emission under $\sigma^+$ excitation. 
    \textbf{b,} $\rho$ under $\sigma^+$ excitation.}
    \label{fig:ext_fig5}
\end{figure*}

\begin{figure*}[htbp]
    \centering
    \includegraphics[width=\textwidth]{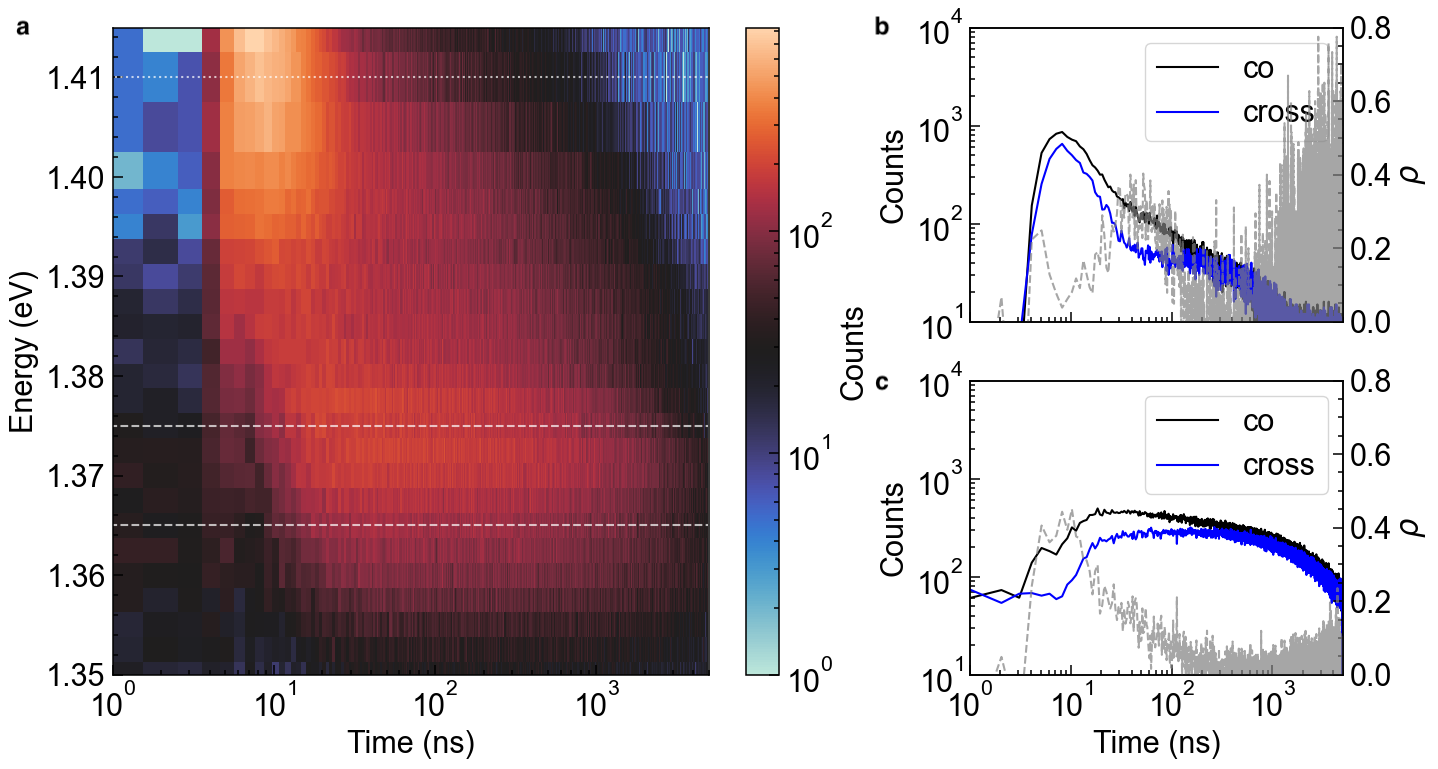}
    \caption{\textbf{Energy resolved IX and IXX traces. a,} Sum of emission under $\sigma^+$ excitation. 
    \textbf{b}, Trace of IXX at \SI{1.41}{\eV} from co(cross)-polarized channel in black (blue), as well as calculated $\rho$. 
    \textbf{c,} Trace of IX integrated between \SI{1.365}{\eV} to \SI{1.375}{\eV} from co(cross)-polarized channel in black (blue), as well as calculated $\rho$. The traces in Figure \ref{fig:fig4}b-d, were captured at a different spot after a new cooldown cycle.}
    \label{fig:ext_fig6}
\end{figure*}

\clearpage

\bibliography{apssamp_used}

\end{document}